\documentclass[aps,pra,preprint,groupedaddress,showkeys]{revtex4}
\usepackage[dvips]{graphicx}
\begin{document}

\pagenumbering{arabic} \pagestyle{myheadings}

{\bf  ELLIPTIC DICHROISM IN ANGULAR DISTRIBUTIONS IN FREE-FREE
TRANSITIONS  IN HYDROGEN}

\vspace*{1.3cm} \noindent
\hspace*{1in} Aurelia Cionga$^1$, Fritz Ehlotzky$^2$, and Gabriela Zloh$^1$
\newline

\noindent
\hspace*{1in}{\ $^1$Institute for Space Science, } \newline
\hspace*{1.15in}{{P.O. Box MG-23, R-76900 Bucharest, Romania}\newline
\hspace*{1in}{\ $^2$Institute for Theoretical Physics, University of
Innsbruck}\newline
\hspace*{1.05in}{\ Technikerstrasse 25, A-6020 Innsbruck, Austria}
\vspace*{1.3cm} }

\noindent
{\bf INTRODUCTION} \vspace*{ 0.6cm }

Dichroism is a well known concept in classical optics where it denotes the
property shown by certain materials of having absorption coefficients which
depend on the state of polarization of the incident light$^1$.
This
 concept has been extended to the case of atomic or molecularinteractions
 with a radiation field. In particular, the notion of ellipticdichroism in
 angular distribution (EDAD) refers to the difference between the
differential cross sections (DCS) of laser assisted signals for {\it left}
and {\it right} elliptically polarized ($EP$) light$^2$.

We discuss here the effect of the photon helicity in laser induced and
inverse bremstrahlung for {\it high energy scattering} of electrons by
hydrogen atoms. We demonstrate that it is possible to find EDAD for high
scattering energies of the electrons if the{\it \ dressing of the atomic
target} by the laser field is taken into account. We consider higher optical
frequencies of the laser field and we restrict ourselves to the use of
moderate
 field intensities. In this case we can employ a hybrid treatment ofthe
 problem$^3$: the interaction between the scattered electronand the
 laser field is described by Volkov solutions, while thelaser-dressing of
 the atomic electron is evaluated within the framework oftime-dependent
 perturbation theory (TDPT). Using this approximation, we candemonstrate
 that EDAD becomes a non-vanishing effect provided second orderTDPT is used
 to describe the dressing of the atomic electron by the ellipticlaser field.
 Moreover, we analyze the role of the virtual transitions betweenthe bound
 and continuum states and we show that these transitions areessential in
 order to be able to predict the existence of EDAD effects.

The basic equations will be presented in section II. In section III we shall
consider in some detail the case of two photon transitions in the weak field
limit. The helicity dependence of the DCS for two photon absorption/emission
by the colliding system in interaction with the $EP$ laser field will be
discussed in
 section IV. 

\vspace*{1.3cm}
\noindent
{\bf BASIC EQUATIONS}

\vspace*{0.6cm} We consider free-free transitions in electron-hydrogen
scattering in the presence of an $EP$ laser field of polarization vector $%
\vec{\varepsilon}$ given by
\begin{equation}
\vec{\varepsilon}=\cos \left( \xi /2\right) \left[ \vec{e}_{{\rm i}}+i\vec{e}%
_{{\rm j}}\tan \left( \xi /2\right) \right] ,  \label{pol}
\end{equation}
where $\xi $ is the ellipticity, $-\pi /2\leq \xi \leq \pi /2$, and $\vec{e}%
_{{\rm i,j}}$ are orthogonal unit vectors in the polarization plane.
We are
 particularly interested to know whether the DCS are sensitive to the{\it helicity} of the $EP$ photons, defined by
$\eta =i\vec{n}\cdot \left( \vec{\varepsilon}
\times {\vec{\varepsilon}}^{\;*}\right) \equiv \sin \xi
$,
with $\vec{n}$ the direction of propagation of the $EP$ laser
beam. Right hand $EP$ has $\eta <0$, it corresponds to $-\pi/2 \leq\xi < 0$.
Left hand $EP$ that has opposite helicity, $\eta >0 $,
corresponds to $0 < \xi \leq \pi/2$. For  optical frequencies we adopt the
dipole approximation, thus the resulting  electric field can be described by
\begin{equation}
\vec{{\cal E}}\left( t\right) =i\;\frac{{\cal E}_0}{2} \vec{\varepsilon}
\exp \left( -i\omega t\right) +{\rm c.c.},  \label{field}
\end{equation}
where the intensity of the laser field is given by $I={\cal E}_0^2$.

We assume that at moderate laser field intensities the interaction between
the laser field and the atomic electron can be described$^3$ by TDPT.
We find it necessary to use {\it second order perturbation theory} and,
following Florescu et al$^4$, the approximate solution for the
ground state
 of an electron bound to a Coulomb potential in the presence ofan $EP$ laser
 field can be written in the form
\begin{equation}
|\Psi _1\left( t\right) >=e^{-i{\rm E}_1t}\left[ |\psi _{1s}>+|\psi
_{1s}^{(1)}>+|\psi _{1s}^{(2)}>\right] .  \label{fun}
\end{equation}
Here $|\psi _{1s}>$ is the unperturbed ground state of the hydrogen atom, of
energy ${\rm E}_1$ and $|\psi _{1s}^{(1),(2)}>$ denote the first and second
order laser field dependent corrections, respectively. We made use of the
published expressions$^{4,5}$ of these corrections.

The scattering electron of kinetic energy $E_k$ and momentum $\vec{k}$ in
interaction
 with the field (\ref{field}) can be described by the well known
Gordon-Volkov solution
\begin{equation}
\chi _{\vec{k}}(\vec{r},t)=\frac 1{(2\pi )^{3/2}}\exp {\left\{ -iE_kt+i\vec{k%
}\cdot \vec{r}-i\vec{k}\cdot \vec{\alpha}(t)\right\} },  \label{vol}
\end{equation}
where $\vec{\alpha}\left( t\right) $ describes the classical oscillation of
the electron in the electric field ${\vec{{\cal E}}}(t)$. The amplitude of
this oscillation is given by $\alpha _0=\sqrt{I}/\omega ^2$. Using Graf's
addition theorem$^6$ of Bessel functions, the Fourier expansion of
the Gordon-Volkov solution (\ref{vol}) leads to a series in terms of
ordinary Bessel functions $J_N$ since one has
\begin{equation}
\exp \left[ -i\vec{k}\cdot \alpha (t)\right] =\exp \left\{ -i{\cal R}_k\sin
\left( \omega t-\phi _k\right) \right\} =\sum_{N=-\infty }^{N=\infty
}J_N\left( {\cal R}_k\right) \exp \left[ -iN\left( \omega t-\phi _k\right)
\right] .
\end{equation}
Following the definitions of the arguments and phases given in Watson's
book,$^6$ we can write
\begin{equation}
{\cal R}_k=\alpha _0\cos (\xi /2)\sqrt{(\vec{k}\cdot {\vec{e}}_i)^{\;2}+(%
\vec{k}\cdot {\vec{e}}_j)^{\;2}\tan ^2(\xi /2)}\equiv
\alpha _0|\vec{k}\cdot \vec{\varepsilon}|
\end{equation}
and
\begin{equation}
\exp (i\phi _k)=\frac{\vec{k}\cdot \vec{\varepsilon}}
{|\vec{k}\cdot\vec{\varepsilon}|}.  \label{exp}
\end{equation}
We recognize that a change of sign of the helicity of the $EP$ photons,
corresponding to the replacement $\vec{\varepsilon}\to \vec{\varepsilon}%
^{\;*}$, will lead to a change in sign of the dynamical phase $\phi _k$.
Therefore, by searching for the signature of helicity in the angular
distributions of the scattered electrons, it will be crucial to look for the
presence of the dynamical phase in the expressions of the DCS.

For high scattering energies, the first order Born approximation in terms of
the interaction potential is reliable. Neglecting exchange effects, this
potential is given by $V(r,R)=-1/r+1/|\vec{r}+\vec{R}|$, where $\vec{R}$
refers to the atomic coordinates. Then, the $S-$matrix element reads
\begin{equation}
S_{fi}^{B1}=-i\int_{-\infty }^{+\infty }dt<{\chi }_{{\vec{k}}_f}(t)\Psi
_1(t)|V|{\chi }_{{\vec{k}}_i}(t)\Psi _1(t)>,  \label{GZ}
\end{equation}
where $\Psi _1(t)$ and ${\chi }_{{\vec{k}}_{i,f}}(t)$ are given by the
dressed atomic state (\ref{fun}) and by the Gordon-Volkov states (\ref{vol}%
), respectively. $\vec{k}_{i(f)}$ represent the initial(final) momenta of
the scattered electron. After Fourier decomposition of the $S$-matrix
element (\ref{GZ}), the DCS for a scattering process involving $N$ laser
photons can be written in the standard form
\begin{equation}
\frac{d\sigma _N}{d\Omega }={(2\pi )}^4\frac{k_f{(N)}}{k_i}|T_N|^2.
\label{sed}
\end{equation}
$N$ is the net number of photons exchanged between the colliding system and
the laser field (\ref{field}), thus the scattered electrons have the final
energy $E_f=E_i+N\omega $. ($N\geq 1$ refers to the absorption and $N\leq -1$
to the emission of laser quanta, while $N=0$ corresponds to the elastic
scattering process.)

In the foregoing equation (\ref{sed}), the nonlinear transition matrix
elements $T_N$, obtained from the $S$-matrix element (\ref{GZ}), have the
following general structure
\begin{equation}
T_N=\exp \left( iN\phi _q\right) \left[ T_N^{(0)}+T_N^{(1)}+T_N^{(2)}\right]
.  \label{tm-cp}
\end{equation}
$\phi _q$ is the dynamical phase defined in (\ref{exp}), referring here to
the momentum transfer $\vec{q}=\vec{k}_i-\vec{k}_f$ of the scattered
electron. The first term in equation (\ref{tm-cp}),
\begin{equation}
T_N^{(0)}=-\left(2\pi \right)^{-2}f_{el}^{B1}J_N({\cal R}_q)\;,
\label{t0}
\end{equation}
would yield the well-known Bunkin-Fedorov formula$^7$.
$f_{el}^{B1}$
 is the amplitude of elastic electron scattering in the firstorder Born
 approximation: $f_{el}^{B1}=2\left( q^2+8\right) /\left(q^2+4\right) ^2$.
 The other two terms in the transition matrix element(\ref{tm-cp}) are
 related to the atomic dressing by the laser field. Theseterms were
 discussed in considerable detail in our precedingwork$^8$. The
 second term, $T_N^{(1)}$, refers to first orderdressing of the atom in
 which case {\it one} of the $N$ photons exchangedbetween the colliding
 system and the radiation field is interacting with thebound electron, while
 the third term $T_N^{(2)}$ refers to second orderdressing and here {\it two}
 of the $N$ photons exchanged during thescattering interact with the atomic
 electron. For an $EP$ field, thesedressing terms are given by
 \begin{equation}
T_N^{(1)} =\frac{\alpha _0\omega }{4\pi ^2q^2} \frac{| \vec{q}\cdot \vec{%
\varepsilon} |}q\;\left[ J_{N-1}({\cal R}_q)\;-J_{N+1}({\cal R}_q)\;\right]
{\cal J}_{1,0,1}\left( q; \omega \right)  \label{t1} \\
\end{equation}
and
\begin{eqnarray}
T_N^{(2)}=\frac{\alpha _0^2\omega ^2}{8\pi ^2q^2}\left\{ J_{N-2}({\cal R}%
_q)\right. \left[ {| \vec{q}\cdot \vec{\varepsilon} |^2}{q^{-2}}{\cal T}_1
\left( q; \omega \right)\right. &+&\left. {\cal T}_2 \left( q; \omega
\right) \cos \xi \;e^{-2i\phi _q}\right]  \nonumber \\
+J_{N+2}({\cal R}_q) \left[ {| \vec{q}\cdot \vec{\varepsilon} |^2} {q^{-2}}
{\cal T}_1 \left( q; \omega \right)\right. &+&\left. {\cal T}_2 \left( q;
\omega \right) \cos \xi \; e^{2i\phi _q}\right]  \nonumber \\
+J_N({\cal R}_q)\left[ {| \vec{q}\cdot \vec{\varepsilon} |^2}{q^{-2}}%
\widetilde{{\cal T}}_1 \left( q; \omega \right)\right. &+&\left. \widetilde{%
{\cal T}}_2 \left( q; \omega \right)\right] \left. {}\right\} .  \label{t2}
\end{eqnarray}
The five radial integrals, denoted by ${\cal J}_{1,0,1}$, ${\cal T}_1$, $%
{\cal T}_2$, $\widetilde{{\cal T}}_1$ and $\widetilde{{\cal T}}_2$ in the
foregoing equations (\ref{t1}) and (\ref{t2}), depend not only on the
absolute value of the momentum transfer $q$ but also on the photon
frequency. For the numerical evaluations performed in the present work, we
used the analytic expressions for the above five radial integrals which are
presented explicitly elsewhere$^{8-10}$.

The transition matrix elements $T_N^{(1)}$ and $T_N^{(2)}$ in (\ref{t1})-(%
\ref{t2}) are written in a form which evidently permits to analyze their
dependence on the dynamical phase $\phi _q$. We recognize immediately that $%
T_N^{(0)}$ and $T_N^{(1)}$ do not depend on the helicity of the photon. On
the contrary, $T_N^{(2)}$ exhibits such an explicit dependence. This
dependence is determined by the phase factors $e^{\pm 2i\phi _q}$ by which $%
{\cal T} _2$ is multiplied in (\ref{t2}). This demonstrates the necessity to
describe target dressing in second order TDPT. In order to stress the
important role of the virtual transitions to the continuum, we shall analyze
small scattering angles. Here the dressing of the target is considerable and
the EDAD effect can be large.

\vspace*{1.3cm}
\noindent

{\bf WEAK FIELD LIMIT }

\vspace*{.6cm} For small arguments of the Bessel functions, {\it i.e}.
either for weak fields at any scattering angle or for moderate fields at
small scattering angles, we can keep the leading terms in (\ref{tm-cp})
only. We discuss in some detail the case of two photon absorption, $N=2$.
The corresponding matrix element is
\begin{equation}
T_2=\frac{\alpha _0^2}{8\pi ^2q^2}\left[ \left(\vec{q} \cdot \vec{\varepsilon%
}\;\right) ^2{\cal A}\left( q;\omega \right) + \cos \xi \; {\cal B}\left(
q;\omega \right)\right] ,  \label{abs}
\end{equation}
where the amplitudes ${\cal A}$ and ${\cal B}$ depend on $q$ and on $\omega $
\begin{eqnarray}
{\cal A}\left( q;\omega \right) &=&-\frac{q^2}{2^2}\left[ f_{el}^{B1}-\frac{%
4\omega }{q^3}{\cal J}_{1,0,1}-\frac{4\omega ^2}{q^4}{\cal T}_1\right] ,
\nonumber \\
{\cal B}\left( q;\omega \right) &=&{\omega ^2}{\cal T}_2.  \label{ab}
\end{eqnarray}

If we consider $N=-2$ ({\it i.e.} two photon emission), then the complex
conjugate of the $EP$ polarization vector ${\vec{\varepsilon}}$ in (\ref{abs}%
) has to be used and the invariant amplitudes for the appropriate value of $%
q $ have to be evaluated since $q=|\vec{k}_i-\vec{k}_f|$ is a function of $N$.

The DCS formula, derived from the transition matrix element (\ref{abs}),
turns out to be
\begin{equation}
\frac{d\sigma _2}{d\Omega } =\alpha _0^4\frac{k_f}{k_i}\frac
1{2^2q^4}\left\{ \mid \vec{q}\cdot \vec{\varepsilon}\mid ^4\mid {\cal A}\mid
^2+\cos^2 \xi \; \mid {\cal B}\mid ^2 + 2\cos \xi \;{\rm Re}\left[ \left(
\vec{q}\cdot \vec{\varepsilon}\right) ^2 {\cal A}{\cal B}^{\;*}\right]
\right\} .  \label{fin}
\end{equation}
This formula is sensitive to a change of helicity only if ${\rm Im}\;{\cal A}%
\neq 0$ and ${\rm Im}\;{\cal B}\neq 0$. This happens if virtual transitions
to continuum states are energetically allowed$^{8-9}$.

EDAD, defined as the difference between the DCS for left hand ($LH$) and
right hand ($RH$) elliptic polarizations, follows from (\ref{fin}) as
\begin{equation}
\Delta _E=-\alpha_0^4 \;\frac{k_f}{k_i} \;
\frac {q_{{\rm i}}q_{{\rm j}}}{2q^4} \;
\sin (2\xi ){\rm Im}\left( {\cal A}^{*}%
{\cal B}\right) \quad {\rm where}\quad q_{{\rm i;j}}=\vec{q}\cdot {\vec{e}}_{%
{\rm i;j}}.  \label{edad}
\end{equation}
$\Delta _E$ depends on the ellipticity $\xi $, its maximum value corresponds
to $\xi =\pi /4$. $\Delta _E$ is also symmetric with respect to the
replacement $\xi \to \pi /2-\xi $.

\vspace*{1.3cm}

\noindent
{\bf RESULTS AND DISCUSSION} \vspace*{ 0.6cm }

We discuss numerical results for EDAD in laser-assisted electron-hydrogen
scattering at high energies of the ingoing particle. We concentrate our
analysis on the cases in which the number of exchanged photons between the
scattering system and the laser field is $N=\pm2$ since here the effects
turn out to be large enough to be accessible to observation. On the basis of
the formalism developed above, we present the DCS evaluated from (\ref{sed})
for a fixed scattering angle $\theta $ as a function of the azimuth
$\varphi $. We also present the dichroism $\Delta _E$ in the same azimuthal
plane. As initial
 energy of the scattered electrons we have taken $E_i=100$eV and we have
 chosen a laser frequency that is close to an atomic resonance,namely $%
 \omega =10$ eV. We show numerical results for the moderate fieldintensity $%
 I=3.51\times 10^{12}$ Wcm$^{-2}$. The initial electron momentum$\vec{k}_i$
 is taken to point along the $z$-axes; the $EP$ laser beampropagates along
 the same axes.

In figure 1 we plot the DCS at $\theta =20^{\circ }$ for two photon
emission ($N$=-2) in panel (a) and
 two photon absorption ($N$=2) in panel(b).  The signals obtained for $LH$ polarization ($ \xi =\pi/4$) are
represented by
 full lines and those obtained for $RH$ polarization ($ \xi=-\pi/4$) by dotted lines. None of
the two axes of the ellipse are symmetryaxes for the DCS, but a change of helicity $LH \leftrightarrow RH$ is
equivalent to a reflection in each of the planes $xOz$ and $yOz$. The
elliptic dichroism is shown in panel (c).
In this geometry
\begin{equation}
q_{{\rm i}}q_{{\rm j}} = k_f^2 \sin^2 \theta
	\sin \varphi \cos \varphi
\end{equation}
and $\Delta _E$ has an
overall $\sin \left( 2\varphi \right) $ dependence that determines its
four-leaved clover pattern.
 The outer clover leaves correspond to absorption,while the inner ones are
 obtained for emission. The signs of the leaves aredifferent for emission
 and absorption, respectively.

The nonlinear signals exhibit also a strong ellipticity dependence, shown in
figure 2, where the DCS for two photon absorption are plotted for $LH$
elliptic polarization using three values of the ellipticity, namely $\xi
=10^{\circ }$, $\xi =45^{\circ }$, and $\xi =80^{\circ }$. It is interesting
to note that although the DCS may be so different, in this geometry they will
always lead to
 a four-leaved clover pattern. According to (\ref{edad}) onlythe magnitude of the clover leaves will
 be modulated by the ellipticity.
Summarizing we find that EDAD in free-free transitions at high projectile
energies in a EP laser field can only be predicted if laser dressing of the
target atom is taken into account in second order TDPT including virtual
bound-continuum transitions irrespective of the scattering configuration
considered. 

\vspace*{1.3cm} \noindent
{\bf Acknowledgments}\newline

This work has been supported by the Jubilee Foundation of the Austrian
National Bank under project number 6211 and by a special research project
for 2000/1 of the Austrian Ministry of Education, Science and Culture. We
also acknowledge financial support by the University of Innsbruck under
reference number 17011/68-00.

\footnotesize
\vspace*{1.3cm} \noindent
{\bf REFERENCES}

\vspace*{.6cm}
\setlength{\parindent}{0cm}
1. M. Born, {\it Optik} (Springer, Berlin, 1981).

2. N. L. Manakov, A. Maquet, S. I. Marmo, V. Veniard and G. Ferrante,
Elliptic dichroism and angular distribution of electrons in two-photon ionization of atoms
{\it J. Phys. B }{\bf 32}, 3747 (1999).

3. F. W. Byron Jr. and C. J. Joachain,
Electron-atom collisions in a strong laser field
{\it J. Phys. B }{\bf 17}, L295 (1984).

4. V. Florescu, A. Halasz, and M. Marinescu,
Second-order corrections to the wave functions of a Coulomb-field
electron in a weak uniform harmonic electric field
 {\it Phys. Rev. A}{\bf 47 }, 394 (1993).

5. V. Florescu and T. Marian,
First-order perturbed wave functions for the hydrogen atom in
a harmonic uniform external electric field,
{\it Phys. Rev. A }{\bf 34}, 4641 (1986).

6. G. N. Watson, {\it Theory of Bessel Functions}, 2nd. Ed.
(University Press, Cambridge, 1962), p. 359.

7. F. V. Bunkin and M. V. Fedorov,  Bremsstrahlung in a strong radiation
field, {\it Zh. Eksp. Theor. Fiz.}
 {\bf 49 }, 1215 (1965) [{\it Sov. Phys.
JETP}, {\bf 22}, 884 (1966)].

8.  A. Cionga, F. Ehlotzky, and G. Zloh,
Electron-atom scattering in a circularly polarized laser field,
{\it Phys. Rev. A} {\bf 61}, 063417 (2000).

9. A. Cionga and V. Florescu,
 One-photon excitation in the e-H collision in the presence of a laser field,
{\it Phys. Rev. A }{\bf 45}, 5282 (1992).

10.  A. Cionga and G. Zloh, unpublished.

\begin{figure}
\includegraphics[width=3.5in,angle=0]{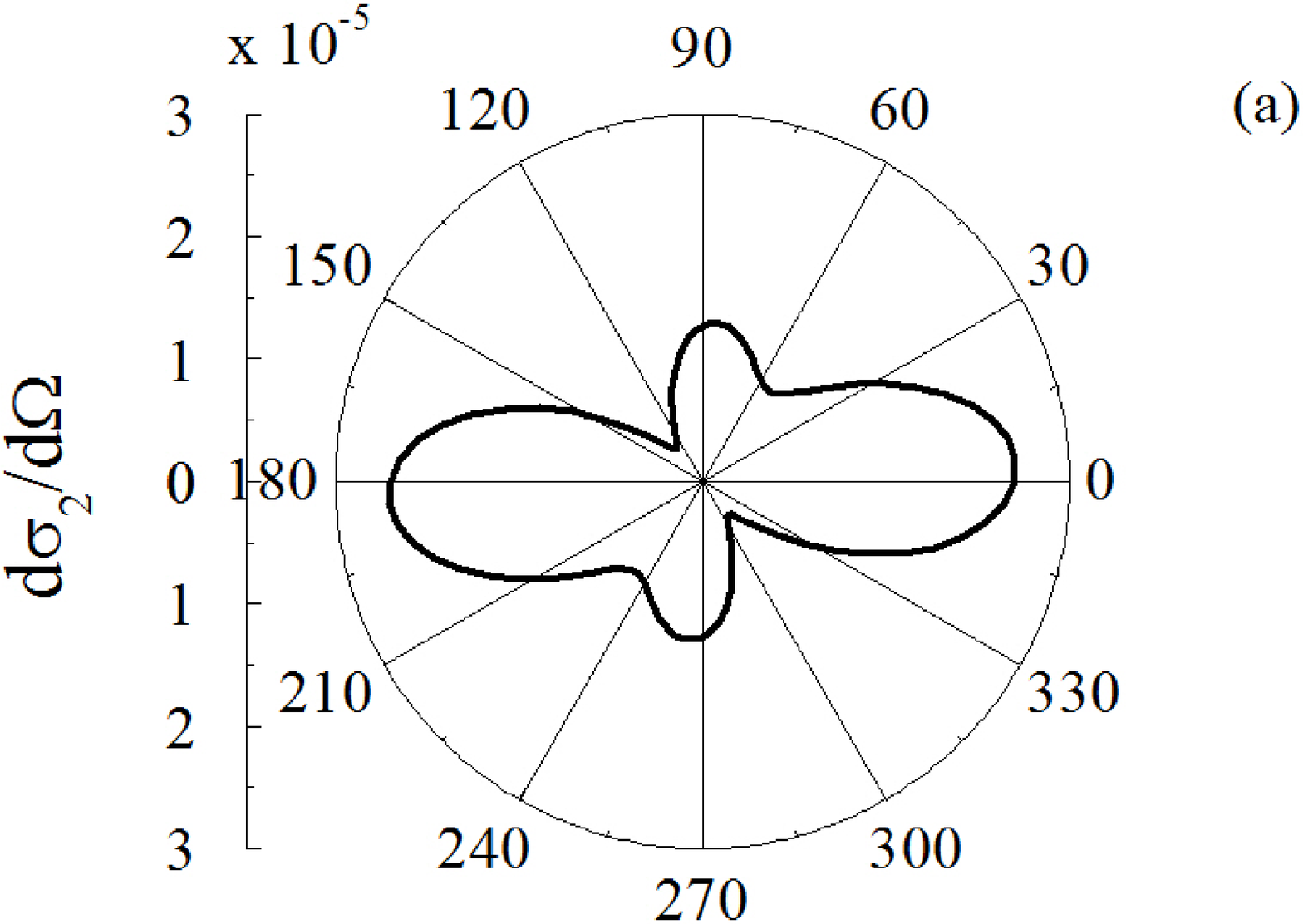}
\includegraphics[width=3.5in,angle=0]{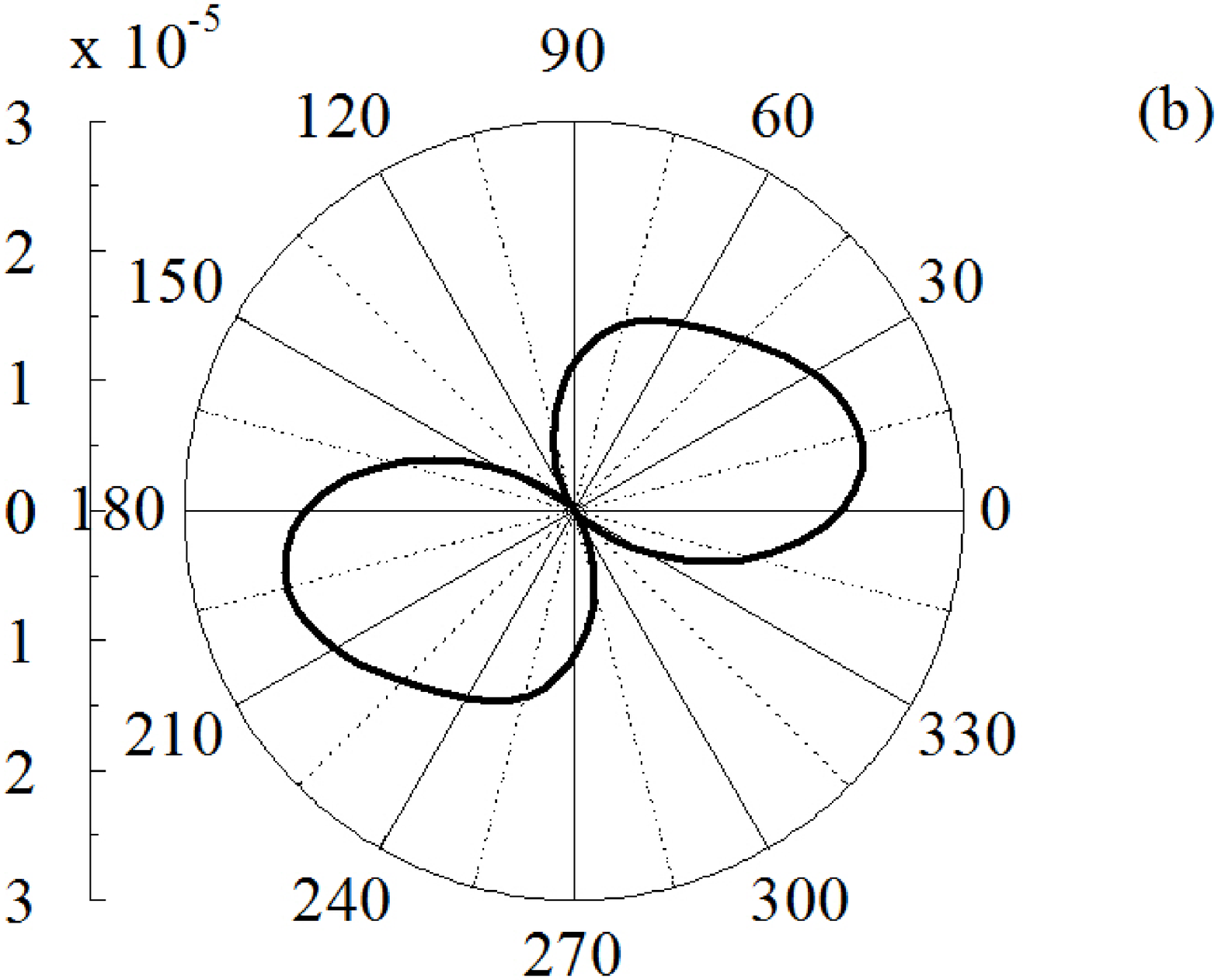}
\includegraphics[width=3.5in,angle=0]{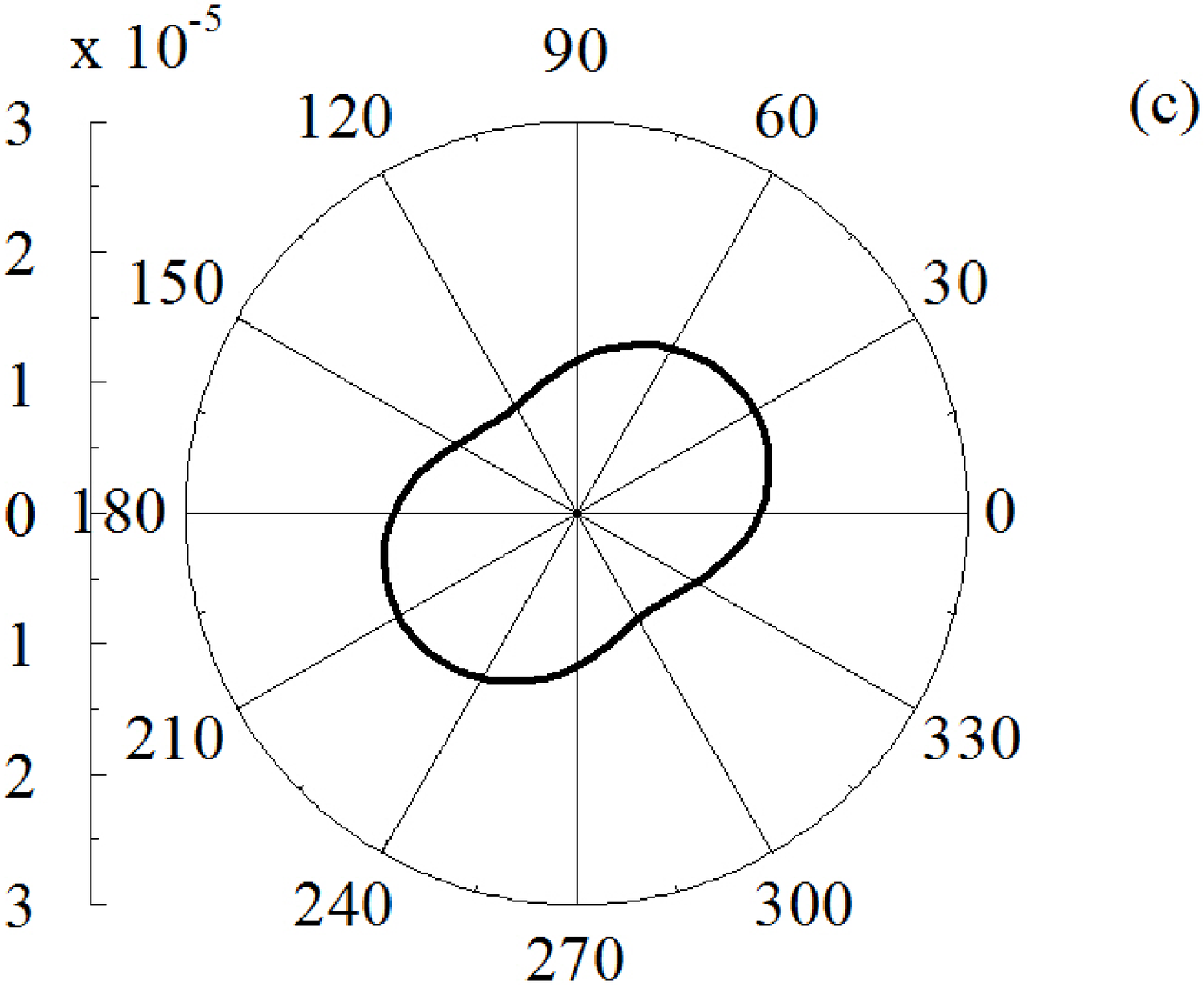}
\centering
\caption{ \footnotesize (a) DCS for two photon emission as a function of the
azimuthal
 angle $\varphi $ at the scattering angle $\theta =20^{\circ }$.Full line is
 used for $LH$ elliptic polarization and dotted line for $RH$elliptic
 polarization. (b) Same as panel (a) but for two photon absorption.In panel
 (c) the dichroism $\Delta _E$ is shown. The outer clover correspondto $N$%
 =2, while the inner one is obtained for $N$=-2. }
\end{figure}

\newpage
 \begin{figure}
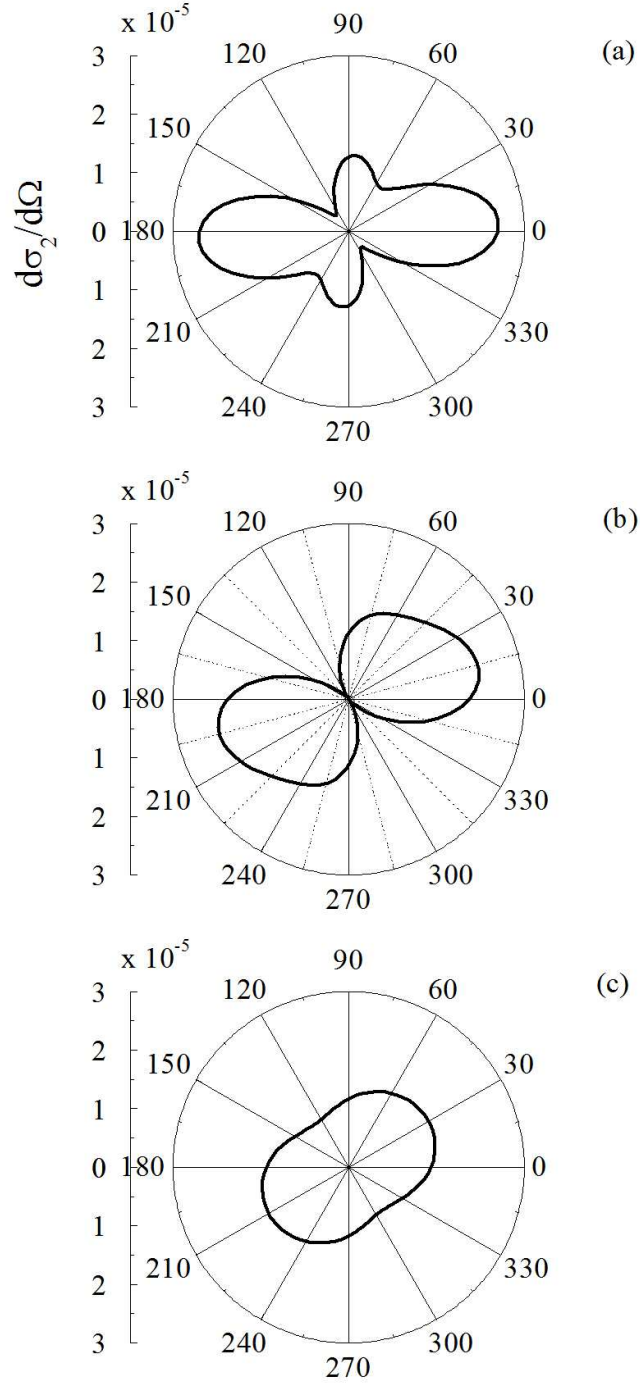

\includegraphics[width=3.5in,angle=0]{Fig3a.eps}
\includegraphics[width=3.5in,angle=0]{Fig3b.eps}
\includegraphics[width=3.5in,angle=0]{Fig3c.eps}
\centering
\caption{\footnotesize (a) DCS for two photon absorption ($N$=2) as a function
of the
 azimuthal angle $\varphi $ for $LH$ helicity and ellipticity $\xi
=10^{\circ }$. The scattering angle is $\theta =10^{\circ }$. The restof the parameters are the same as in
 figure 1(b). (b) same as in panel2(a) but for $\xi =45^{\circ }$. (c) same as in panel
2(a) but for $\xi =80^{\circ }$. }
 \end{figure}

\end{document}